\documentclass[a4paper,11pt]{article}

\usepackage{amsmath}
\usepackage{graphicx}
\usepackage{cite}

\begin{document}

\newcommand{\hatb}[1]{{\hat{\bf #1}}}
\renewcommand{\vec}[1]{\mbox{${\bf #1}$}}
\renewcommand{\v}[1]{{{\bf #1}}}
\newcommand{\dyad}[1]{\mbox{$\overline{ \v{#1} }$}}
\newcommand{\romvec}[1] {\mbox{\boldmath $#1$}}
\newcommand{\lagr}{\mbox{\boldmath $\lambda$}}
\newcommand{\intsol}{\mbox{\boldmath $\phi$}}
\newcommand{\rhov}{\mbox{\boldmath $\rho$}}
\newcommand{\tr}{{\dagger}}
\newcommand{\pd}[2]{\frac{\partial #1}{\partial #2}}
\newcommand{\dd}[2]{\frac{\text{d} #1}{\text{d} #2}}
\newcommand{\od}[2]{\frac{d #1}{d #2}}

\def\dyad #1{\v{\overline{#1}}}
\def\vE{\v E}
\def\vH{\v H}
\def\vD{\v D}
\def\vB{\v B}

\newcommand{\scriptl}{$\mathbb{L}$}

\def\dyad #1{\v{\overline{#1}}}
\def\dyadg #1{{\boldsymbol{\overline{#1}}}}

\def\be{\begin{equation} }
\def\ee{\end{equation} }

\def\domain #1{D(\mathbb{#1})}
\def\range #1{R(\mathbb{#1})}

\def\vromega{(\v r, \omega)}

\newcount\timehh\newcount\timemm
\timehh=\time \divide\timehh by 60 \timemm=\time \count255=\timehh
\multiply\count255 by -60 \advance\timemm by \count255
\def\draftheader{\slshape\today\ at
\ifnum\timehh<10 0\fi\number\timehh\,:\,\ifnum\timemm<10 0\fi\number\timemm}%

\newcommand{\vg} [1]{\mbox{\boldmath $#1$}}

\def\norm #1{\left\| {#1} \right\|}

\def\Krylov#1{\mathcal K^{#1}\left(\dyad A,\v r_0\right)}

\newcommand{\jeq}{\mathbf J_{eq}}
\newcommand{\rbar}{(\mathbf r)}
\newcommand{\rbarp}{(\mathbf r')}
\newcommand{\bg}{\mathbf{\overline G}}
\newcommand{\prs}{(\mathbf r, \mathbf r')}
\newcommand{\meq}{\mathbf M_{eq}}
\newcommand{\hinc}{\mathbf H_{inc}}
\newcommand{\einc}{\mathbf E_{inc}}

\def\tdot{\hat t\cdot}
\def\Itd{\dyad I_t\cdot}
\def\It{\dyad I_t}

\newcommand{\js}{\mathbf J_{s}}
\newcommand{\bt}{\mathbf T}
\newcommand{\bbf}{\mathbf F}
\newcommand{\bmu}{\boldsymbol{\overline{\mu}}}
\newcommand{\beps}{\boldsymbol{\overline{\epsilon}}}
\newcommand{\bbtau}{\boldsymbol{\overline{\tau}}}
\newcommand{\bbI}{{\bf\overline{I}}}
\newcommand{\bbg}{\mathbf{\overline G}}
\newcommand{\bbi}{{\bf\overline I}}

\newcommand{\jays}{\mathbf J_s}
\newcommand{\ms}{\mathbf M_s}

\newcommand{\eneg}{e^{-ik_1z}}
\newcommand{\epos}{e^{ik_1z}}
\newcommand{\ein}{e^{in\phi}}

\def\Deltamur{\Delta\bmu_r}
\def\Deltaepsr{\Delta\beps_r}

\newcommand{\bbarz}{\mathbf{\overline Z}}
\newcommand{\bbarp}{\mathbf{\overline P}}
\newcommand{\bbark}{\mathbf{\overline K}}
\newcommand{\bbara}{\mathbf{\overline A}}
\newcommand{\bbarb}{\mathbf{\overline B}}
\newcommand{\bbard}{\mathbf{\overline D}}
\newcommand{\kayq}{{\mathbf{\overline K}}^{-1}\cdot\mathbf Q}
\def\MHE{electromagnetic equations}

\renewcommand{\v}[1]{{{\bf #1}}}
\def\dyad #1{\v{\overline{#1}}}
\def\braket#1#2{\langle{#1|#2}\rangle}
\def\bra#1{\langle{#1}|}
\def\ket#1{|{#1}\rangle}

\def\gTM{g^{\textrm{\scriptsize TM}}}
\def\gTE{g^{\textrm{\scriptsize TE}}}
\def\gTEs{g_s^{\textrm{\scriptsize TE}}}
\def\gTMr{g^{\textrm{\scriptsize TM}}\left(\v{r},\v{r}'\right)}
\def\gTEr{g^{\textrm{\scriptsize TE}}\left(\v{r},\v{r}'\right)}
\def\gTEsr{g_s^{\textrm{\scriptsize TE}}\left(\v{r},\v{r}'\right)}

\def\gTMS{g^{\textrm{\scriptsize TM},S}}
\def\gTES{g^{\textrm{\scriptsize TE},S}}
\def\gTEsS{g_s^{\textrm{\scriptsize TE},S}}
\def\gTMSr{g^{\textrm{\scriptsize TM},S}\left(\v{r},\v{r}'\right)}
\def\gTESr{g^{\textrm{\scriptsize TE},S}\left(\v{r},\v{r}'\right)}
\def\gTEsSr{g_s^{\textrm{\scriptsize TE},S}\left(\v{r},\v{r}'\right)}

\def\DG{\dyad G}
\def\curl{\nabla\times}
\def\curlp{\nabla'\times}
\def\div{\nabla\cdot}
\def\divp{\nabla'\cdot}
\def\grad{\nabla}
\def\gradp{\nabla'}
\def\curlDG{\curl\DG}

\def\TMS{{{\rm TM},S}}
\def\TES{{{\rm TE},S}}

\def\opL{\mathcal L}
\def\opK{\mathcal K}

\def\kmns{k_{mn}^2}
\def\pr{(\v r)}
\def\prp{(\v r')}
\def\prs{(\vec{r}, \vec{r}')}

\def\DGLM{dyadic Green's function for layered media}

\def\eeqa{\end{eqnarray}}
\renewcommand{\v}[1]{{{\bf #1}}}

\def\unitnormal{\hat{n}}

\def\MHE{electromagnetic equations}
\def\LVS{linear vector space}
\def\IPS{inner product space}
\def\CEM{computational electromagnetics}
\def\be{\begin{equation} }
\def\ee{\end{equation} }
\def\cm{classical mechanics}
\def\qm{quantum mechanics}
\def\EOM{equation of motion}
\def\EsOM{equations of motion}
\def\QM{Quantum mechanics}
\def\SE{Schr\"{o}dinger equation}
\def\Schr{Schr\"{o}dinger}
\def\tr{\text{tr}}
\def\Tr{\text{Tr}}
\def\EMS{effective mass Schr\"odinger}
\def\rep{representation}
\def\EM{electromagnetic}
\def\PBC{periodic boundary condition}
\def\LHS{left-hand side}
\def\RHS{right-hand side}


%

\def\inf{\infty}
\def\intinf{\int\limits_{-\inf}^{\quad\inf}}
\def\intsinf{\int\limits_{0}^{\quad\inf}}
\def\iintinf{\iint\limits_{-\inf}^{\qquad\inf}}
\def\pd #1#2{\frac {\partial #1}{\partial #2}}
\def\krho{k_\rho}
\def\krhoo{k_{\rho 0}}
\def\hg{\hat g}
\def\today{\ifcase\month\or January\or February\or March\or April\or
May\or June\or July\or August\or September\or October\or November\or
December \fi\space\number\day, \number\year}
\def\inf{\infty}%
\def\dbarL{\bar{\bar {L}}}%
\def\degree{$\,^o$ }
\def\bE{{\bf E}}%
\def\tE{\tilde E}
\def\bJ{{\bf J}}%
\def\vr{{\bf r}}%
\def\dG{\bar{\bf G}_o (\vr ,\vr'')}%
\def\sg{g_o(\vr,\vr')}
\def\dbar#1{\bar{\bar{#1}}}%
\def\endeq{\eqno{\ENUM}}
\def\krho{k_\rho}
\def\~{\tilde}
\def\^{\hat}
\def\pd#1#2{\frac{\partial#1}{\partial#2}}

\def\R{\Re e}
\def\I{\Im m}
\def\eqn#1{\eqno{\text{#1}}}
\def\cal{\fam2 }
\def\lover#1{\buildrel \leftarrow \over {#1}}
\def\PVint{{-}\hskip -13.pt\int\limits}
\def\pvint{{-}\hskip -10pt\int\limits}
\def\m{\multline }
\def\em{\endmultline }

\def\itemitems{\hangindent2\parindent \textindent}
\def\items{\hangindent\parindent \textindent}

\def\MTT{{\it IEEE Trans. Microwave Theory Tech.}  MTT-}
\def\AP{{\it IEEE Trans. Antennas Propagat.}  AP-}
\def\GE{{\it IEEE Trans. Geosci. Remote Sensing}  GE-}
\def\RS{{\it Radio Sci.}  }
\def\EL{{\it Electron. Lett.}  }
\def\ritem{\noindent\hangindent=1.pc\hangafter=1}
\def\eitem#1{\par\noindent\hangindent\parindent\hangafter=1
    \hbox {\bf #1\enspace}\ignorespaces}
\def\q{\quad}
\def\qq{\qquad}
\def\qqq{\qq\q}
\def\qqqq{\qqq\q}

\def\three{\text{$\text{I}\text{I}\text{I}$}}
\def\tinf{\text{\it inf\,}}
\def\math#1{\vphantom{\left(#1\right)}#1}
\def\sqrts#1{ {\sqrt{#1}\,} }
\def\sgn{\,\text{\rm sgn}}

\def\bul{$\bullet$\hskip 1pc}
\def\nl{\vskip \baselineskip}
\def\bitem{\item{\bul}}
\def\subitem#1{\hbox{\hskip 2pc\vtop{\hsize 5.5in\parindent 0pt #1}
\hfil}}
\def\cl#1{\centerline{#1}}
\def\ub#1{\underbar{#1}}
\def\ul#1{\underline{#1}}
\def\Cal#1{{\cal #1}}
\def\Real{\Re e~}
\def\Imag{\Im m~}

\def\Div{\nabla\cdot}
\def\Curl{\nabla\times}
\def\Grad{\nabla}

\begin{center}
{\LARGE \bfseries Vector Potential Electromagnetic Theory with
Generalized Gauge for Inhomogeneous Anisotropic Media}
\end{center}

\vspace{1mm}

\begin{center}
{\small \noindent W. C. Chew}\footnote{U of Illinois, Urbana-Champaign; visiting professor, HKU.}\\
\draftheader
\end{center}

\vspace{5mm}


\def\EH{$\v E$-$\v H$}

\abstract{ Vector and scalar potential formulation is valid from
quantum theory to classical electromagnetics.  The rapid development
in quantum optics calls for electromagnetic solutions that straddle
quantum physics as well as classical physics.  The vector potential
formulation is a good candidate to bridge these two regimes. Hence,
there is a need to generalize this formulation to inhomogeneous
media. A generalized gauge is suggested for solving electromagnetic
problems in inhomogenous media which can be extended to the
anistropic case. The advantages of the resulting equations are their
absence of low-frequency catastrophe. Hence, usual differential
equation solvers can be used to solve them over multi-scale and
broad bandwidth.  It is shown that the interface boundary conditions
from the resulting equations reduce to those of classical Maxwell's
equations.  Also, classical Green's theorem can be extended to such
a formulation, resulting in similar extinction theorem, and surface
integral equation formulation for surface scatterers.  The integral
equations also do not exhibit low-frequency catastrophe as well as
frequency imbalance as observed in the classical formulation using
\EH\ fields.  The matrix representation of the integral equation for
a PEC scatterer is given.
 }

\section{Introduction}

Electromagnetic theory has been guided by Maxwell's equations for
150 years now \cite{Maxwell}.  The formulation of electromagnetic
theory based on $\v E$, $\v H$, $\v D$, and $\v B$, simplified by
Heaviside \cite{Heaviside}, offers physical insight that results in
the development of myriads of electromagnetic-related technologies.
However, there are certain situations where the $\v E$-$\v H$
formulation is not ideal.  This is in the realm of quantum mechanics
where the $\v A$-$\Phi$ formulation is needed. In certain situations
where $\v E$-$\v H$ are zero, but $\v A$ is not zero, and yet, the
effect of $\v A$ is felt in \qm.  This is true of the Aharonov-Bohm
effect \cite{Aharonov,Gasiorowicz}. Moreover, the quantization of
electromagnetic field can be done more expediently with the vector
potential $\v A$ rather than $\v E$ and $\v H$. More importantly,
when the electromagnetic effect needs to be incorporated in \SE,
vector and scalar potentials are used.  This will be important in
many quantum optics studies
\cite{Tannoudji,Mandel,Scully,Loudon,Gerry,Fox,Chiao}.

Normally, electromagnetic equations formulated in terms of $\v
E$-$\v H$ have low-frequency breakdown or catastrophe.  Hence, many
numerical methods based on \EH\ formulation are unstable at low
frequencies or long wavelength. Therefore, the \EH\ formulation is
not truly multi-scale, as it has catastrophe when the dimension of
objects are much smaller than the wavelength. Different formulations
using tree-cotree, or loop-tree decomposition
\cite{Cendes,Jin,Wilton,Vecchi,ZhaoChew}, have to be sought when the
frequency is low or the wavelength is long.
Due to the low-frequency catastrophe encountered by \EH\
formulation, the vector potential formulation has been very popular
for solving low frequency problems
\cite{Chawla,Geselowitz,Demerdash,Biro,MacNeal,Dyczij-Edlinger,Dyczij-Edlinger2,Flaviis,Biro2,Dular,Cangel}.

This work will arrive at a general theory of vector potential
formulation for inhomogeneous anisotropic media, together with the
pertinent integral equations.  This vector potential formulation
does not have apparent low-frequency catastrophe of the \EH\
formulation and it is truly multi-scale.  It can be shown that with
the proper gauge, which is the extension of the simple Lorentz gauge
to inhomogeneous anisotropic media, the scalar potential equation is
decoupled from the vector potential equation.

\def\ME{Maxwell's equations}

\section{Pertinent Equations--Inhomogeneous Isotropic Case}

The vector potential formulation for homogeneous medium has been
described in most text books \cite{Jackson,Harrington,Kong,Balanis}.
 We derive the pertinent equations for the inhomogeneous
isotropic medium case first.  To this end, we begin with the
Maxwell's equations:
\begin{align}\label{e1}
\nabla  \times {\bf{E}} &=  - {\partial _t}{\bf{B}},\\\label{e2a}
\nabla \times {\bf{H}} &= {\partial _t}{\bf{D}}{\rm{ + }}{\bf{J}},
\\\label{e2} \nabla \cdot{\bf{B}} &= 0,
\\\nabla
\cdot {\bf{D}} &= \rho.\label{e2b}
\end{align}
From the above we let
\begin{align}\label{e3}
{\bf{B}} &= \nabla  \times {\bf{A}},\\
{\bf{E}} &=  - {\partial _t}{\bf{A}} - \nabla \Phi\label{e3a}
\end{align}
so that the first and third of four \ME\ are satisfied. Then, using
$\v D=\varepsilon \v E$ for isotropic, non-dispersive, inhomogeneous
media, we obtain that
\begin{equation}\label{e4}
- {\partial _t}\nabla \cdot\varepsilon {\bf{A}} - \nabla \cdot\varepsilon \nabla \Phi  = \rho ,
\end{equation}
\begin{equation}\label{e5}
\nabla  \times {\mu ^{ - 1}}\nabla  \times {\bf{A}} =  - \varepsilon \partial _t^2{\bf{A}} - \varepsilon {\partial _t}\nabla \Phi  + {\bf{J}}.
\end{equation}
For homogeneous medium, the above reduce to
\begin{equation}\label{e6}
- {\partial _t}\nabla \cdot{\bf{A}} - {\nabla ^2}\Phi  = \rho /\varepsilon ,
\end{equation}
\begin{equation}\label{e7}
{\mu ^{ - 1}}\left( {\nabla \nabla  \cdot {\bf{A}} - {\nabla ^2}{\bf{A}}} \right) =  - \varepsilon \partial _t^2{\bf{A}} - \varepsilon {\partial _t}\nabla \Phi  + {\bf{J}}.
\end{equation}
By using the simple Lorentz gauge
\begin{equation}\label{e8}
\nabla  \cdot {\bf{A}} =  - \mu \varepsilon {\partial _t}\Phi
\end{equation}
we have the usual
\begin{equation}\label{e9}
{\nabla ^2}\Phi  - \mu \varepsilon \partial _t^2\Phi  =  - \rho /\varepsilon ,
\end{equation}
\begin{equation}\label{e10}
{\nabla ^2}{\bf{A}} - \mu \varepsilon \partial _t^2{\bf{A}} =  - \mu {\bf{J}}
\end{equation}
Lorentz gauge is preferred because it treats space and time on the
same footing as in special relativity \cite{Jackson}.

For inhomogeneous media, we can choose the generalized Lorentz
gauge. This gauge has been suggested previously, for example in
\cite{Flaviis}.
\begin{equation}\label{e11}
{\varepsilon ^{ - 1}}\nabla  \cdot \varepsilon {\bf{A}} =  - \mu \varepsilon {\partial _t}\Phi.
\end{equation}
However, we can decouple \eqref{e4} and \eqref{e5} with an even more
generalized gauge, namely
\begin{equation}\label{e11a}
\Div \varepsilon\v A =- \chi \partial_t \Phi
\end{equation}
Then we get from \eqref{e4} and \eqref{e5} that
\begin{equation}\label{e12}
\nabla \cdot\varepsilon \nabla \Phi  - \chi
\partial _t^2\Phi  =  - \rho  ,
\end{equation}
\begin{equation}\label{e13}
- \nabla  \times {\mu ^{ - 1}}\nabla  \times {\bf{A}} - \varepsilon
\partial _t^2{\bf{A}} +  \varepsilon \nabla \chi^{-1}\nabla \cdot\varepsilon {\bf{A}}
=  - {\bf{J}}.
\end{equation}
It is to be noted that \eqref{e12} can be derived from \eqref{e13}
by taking the divergence of \eqref{e13} and then using the
generalized gauge and the charge continuity equation that $\Div \v
J=-\partial_t \rho$.  In gneral, we can choose
\begin{equation}\label{chi:eq}
\chi=\alpha\varepsilon^2\mu
\end{equation}
where $\alpha$ can be a function of position.  When $\alpha=1$, it
reduces to the generalized Lorentz gauge used in \eqref{e11}.

 For homogeneous medium, \eqref{e12} and \eqref{e13} reduce to
\eqref{e9} and \eqref{e10} when we choose $\alpha=1$ in
\eqref{chi:eq}, which is the case of the simple Lorentz gauge.
Unlike the vector wave equations for inhomogeneous electromagnetic
fields, the above do not have apparent low-frequency breakdown when
$\partial _t=0$. Hence, the above equations can be used for
electrodynamics as well as electrostatics when the wavelength tends
to infinity.

We can rewrite the above as a sequence of three equations, namely,
\begin{align}
\nabla \cdot\varepsilon {\bf{A}}&=-\chi\partial_t \Phi\label{eq17}\\
\nabla\times\v A&=\mu\v H\label{eq18}\\
\nabla\times\v H+\varepsilon\partial_t^2\v
A+\varepsilon\nabla\partial_t\Phi&=\v J\label{eq19}
\end{align}
The last equation can be rewritten as
\begin{align}\label{eq20}
\nabla\times\v H-\varepsilon\partial_t(-\partial_t\v
A-\nabla\Phi)=\v J
\end{align}
which is the same as solving Ampere's law.  Hence, solving
\eqref{e13} is similar to solving Maxwell's equations.

It is to be noted that \eqref{e13} resembles the elastic wave
equation in solids where both longitudinal and transverse waves can
exist \cite{ChewTongHu}.  Furthermore, these two waves can have
different velocities in a homogeneous medium if $\alpha\ne 1$ in
\eqref{chi:eq}.  The longitudinal wave has the same velocity as the
scalar potential, which is $1/\sqrt{\alpha\mu\varepsilon}$, while
the transverse wave has the velocity of light, or
$1/\sqrt{\mu\varepsilon}$. If we choose $\alpha=0$, or $\chi=0$, we
have the Coulomb gauge where the scalar potential has infinite
velocity.

\section{Boundary Conditions for the Potentials}

\begin{figure}[h]
\begin{center}
\noindent
  \includegraphics[width=3in]{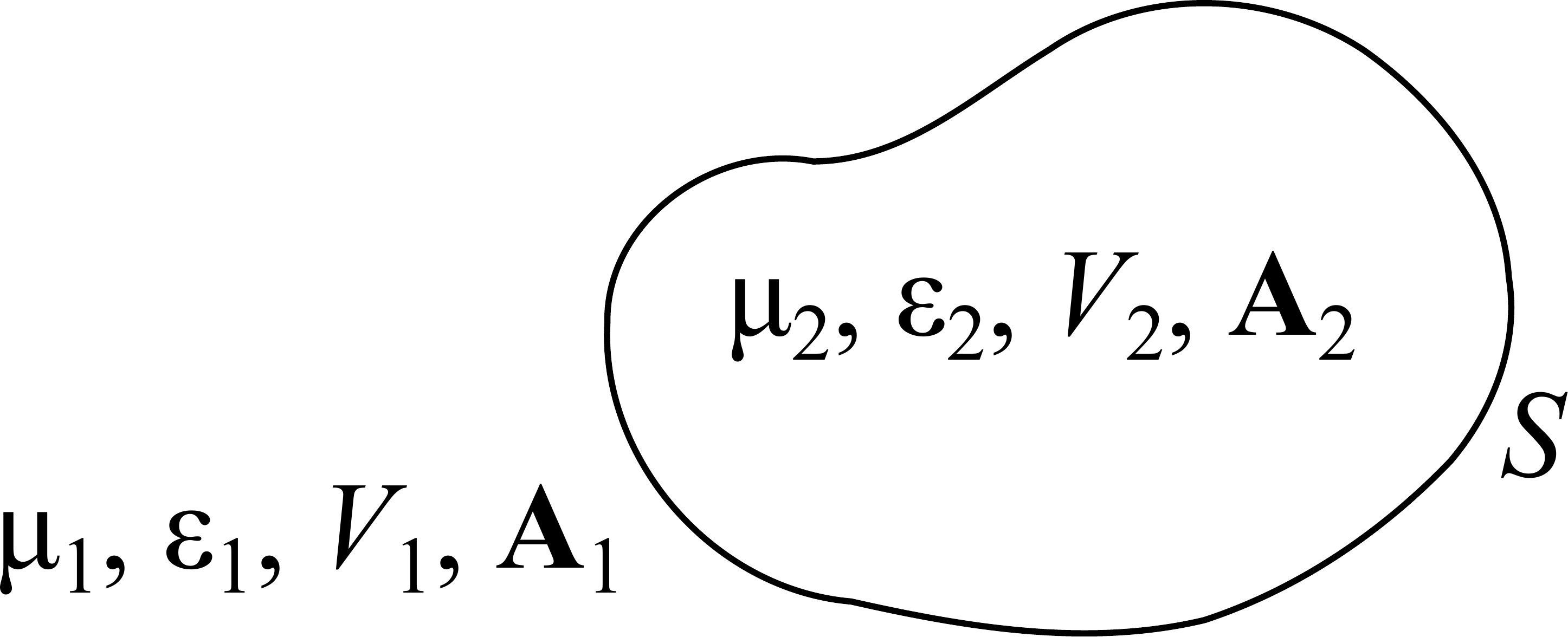}\label{Fig1}
\end{center}
\caption{\bf The boundary conditions for the interface between two
piecewise homogeneous regions.}
\end{figure}

Many problems can be modeled with piecewise homogeneous medium. In
case, the solutions can be sought in each of the piecewise
homogenous region, and then sewn together using boundary conditions.
The above equations, \eqref{e12} and \eqref{e13}, are the governing
equations for the scalar and vector potentials $\Phi$ and $\v A$ for
inhomogeneous media. The boundary conditions at the interface of two
homogeneous media are also embedded in these equations.
%
%
By eyeballing equation \eqref{e13}, we see that $\nabla  \times
{\bf{A}}$ must be finite at an interface. This induces the boundary
condition that
\begin{equation}\label{2:e2}
\hat n \times {{\bf{A}}_1} = \hat n \times {{\bf{A}}_2}
\end{equation}
across an interface. Assuming that $\v J$ is finite, we also have
\begin{equation}\label{2:e3}
\hat n \times \frac{1}{{{\mu _1}}}\nabla  \times {{\bf{A}}_1} = \hat
n \times \frac{1}{{{\mu _2}}}\nabla  \times {{\bf{A}}_2}.
\end{equation}
The above is equivalent to
\begin{equation}\label{2:e4}
\hat n \times {{\bf{H}}_1} = \hat n \times {{\bf{H}}_2}
\end{equation}
at an interface.  When a surface current sheet is present, we have
to augment the above with the current sheet as is done in standard
\EM\ boundary conditions.
Furthermore, due to the finiteness of $\nabla \cdot\varepsilon
{\bf{A}}$ at an interface, it is necessary that
\begin{equation}\label{2:e5}
\hat n \cdot {\varepsilon _1}{{\bf{A}}_1} = \hat n \cdot
{\varepsilon _2}{{\bf{A}}_2}.
\end{equation}
It can be shown that if a surface dipole layer exists at an
interface, we will have to augment the above with the correct
discontinuity or jump condition.  A surface current with a normal
component to the surface will constitute a surface dipole layer.

By the same token, we can eyeball the scalar potential equation
\eqref{e12}, and notice that
\begin{equation}\label{2:e6}
\hat n \cdot {\varepsilon _1}\nabla {\Phi _1} = \hat n \cdot
{\varepsilon _2}\nabla {\Phi _2}.
\end{equation}
The above will be augmented with the necessary jump or discontinuity
condition if a surface charge layer exists at an interface.
Equations (\ref{2:e5}) and (\ref{2:e6}) together mean that
\begin{equation}\label{2:e7}
\hat n \cdot {\varepsilon _1}{{\bf{E}}_1} = \hat n \cdot
{\varepsilon _2}{{\bf{E}}_2}.
\end{equation}
where we have noted that ${\bf{E}}{\rm{ = }} - {\partial _t}{\bf{A}}
- \nabla \Phi$ from \eqref{e3a}.  This is the usual boundary
condition for the normal component of the electric field.

Equation \eqref{e12} also implies that
\begin{equation}\label{2:e8}
{\Phi _1} = {\Phi _2},
\end{equation}
or
\begin{equation}\label{2:e9}
\hat n \times \nabla {\Phi _1} = \hat n \times \nabla {\Phi _2}.
\end{equation}
Equations (\ref{2:e2}) and (\ref{2:e9}) imply that
\begin{equation}\label{2:e10}
\hat n \times {{\bf{E}}_1} = \hat n \times {{\bf{E}}_2}.
\end{equation}
This is the normal boundary condition for the tangential component
of the electric field.

If ${\varepsilon _2}$ is a perfect electric conductor (PEC),
${\varepsilon _2} \to \infty$. From (\ref{e13}), it implies that
${{\bf{A}}_2} = 0$, if $\omega  \ne 0$ or $\partial_t\ne 0$. Then
(\ref{2:e2}) for a PEC surface becomes
\begin{equation}\label{2:e11}
\hat n \times {{\bf{A}}_1} = 0.
\end{equation}
Also, by eyeballing \eqref{e12}, we see that for a PEC, $\Phi_2=0$.
This together with \eqref{2:e8}, \eqref{2:e9}, and \eqref{2:e11}
imply that $\hat n \times {{\bf{E}}_1}=0$ on a PEC surface. For a
perfect magnetic conductor (PMC), ${\mu _2} \to \infty$, from
(\ref{2:e3}) and (\ref{2:e4})
\begin{equation}\label{2:e12}
\hat n \times {{\bf{H}}_1} = 0.
\end{equation}

When $\omega=0$ or $\partial_t=0$, $\v A$ does not contribute to $\v
E$.  But from \eqref{2:e6}, when ${\varepsilon _2} \to \infty$, we
deduce that $\hat n\cdot \nabla \Phi_2=0$, implying that
$\Phi_2=\rm{constant}$ for  $\v r_2\in V_2$ for arbitrary $S$.
Hence, from \eqref{2:e8} and \eqref{2:e9}, $\hat n \times
{{\bf{E}}_1}=0$ on a PEC surface even when $\omega=0$.

\section{General Anisotropic Media Case}

For inhomogeneous, dispersionless, anisotropic media, the
generalized gauge becomes
\begin{equation}\label{eq16}
\nabla \cdot \dyadg{\varepsilon} \cdot \v{A} = - \chi\partial_t \Phi
\end{equation}
In the above, $\chi$ is arbitrary, but we can choose
\begin{equation}
\chi=\alpha|\dyadg{\varepsilon} \cdot \dyadg{\mu} \cdot
\dyadg{\varepsilon} |
\end{equation}
where the vertical bar means determinant. When the medium is
inhomogeneous and isotropic, the above gauge reduces to the
generalized gauge previously discussed.   When $\alpha=1$, the above
reduces to the generalized Lorentz gauge for inhomogeneous isotropic
medium. In general, \eqref{e12} and \eqref{e13} become
\begin{equation}\label{eq14}
\nabla \cdot \dyadg{\varepsilon} \cdot \nabla \Phi - \chi
\partial_t^2 \Phi = -\rho
\end{equation}
\begin{equation}\label{eq15}
\nabla \times \dyadg{\mu}^{-1} \nabla \times \v{A} +
\dyadg{\varepsilon} \cdot \partial^2_t \v{A} - \dyadg{\varepsilon}
\cdot \nabla \chi ^{-1}  \nabla\cdot \dyadg{\varepsilon} \cdot \v{A}
= \v{J}
\end{equation}
The above can be rewritten in the manner of \eqref{eq17} to
\eqref{eq20}, showing that solving the above is the same as solving
the original Maxwell's equations.
The boundary condition \eqref{2:e2} remains the same.  Boundary
condition \eqref{2:e3} becomes
\begin{equation}
\hat{n} \times \dyadg{\mu}_1^{-1} \cdot \nabla \times \v{A}_1 =
\hat{n} \times \dyadg{\mu}_2^{-1} \cdot \nabla \times \v{A}_2
\end{equation}
and boundary condition \eqref{2:e4} remains the same.  Similarly,
boundary condition \eqref{2:e5} becomes
\begin{equation}\label{2:e5a}
\hat n \cdot \dyadg{\varepsilon} _1\cdot{{\bf{A}}_1} = \hat n \cdot
\dyadg{\varepsilon} _2\cdot{{\bf{A}}_2}.
\end{equation}
The boundary condition \eqref{2:e6} becomes
\begin{equation}
\hat{n} \cdot \dyadg{\varepsilon }_1 \cdot \nabla \Phi_1 = \hat{n}
\cdot \dyadg{\varepsilon}_2 \cdot \nabla \Phi_2
\end{equation}
while the other boundary conditions, similar to the isotropic case,
can be similarly derived.

\section{Green's Theorem---Time Harmonic Case}

\begin{figure}[h]
\begin{center}
\noindent
  \includegraphics[width=3in]{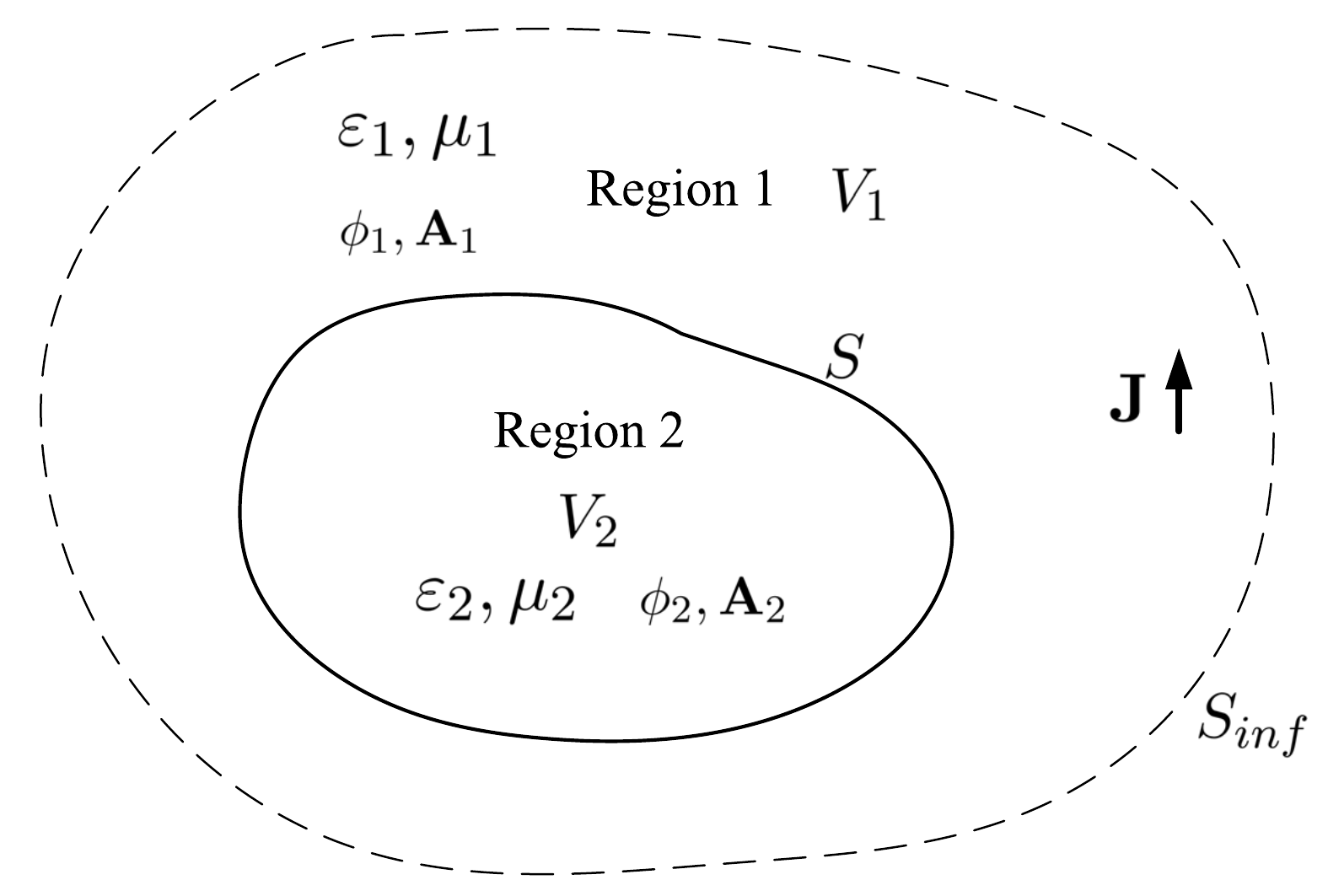}\label{Fig2}
\end{center}
\caption{\bf The figure used to derive the Green's theorem for
vector potential equation.}
\end{figure}

As mentioned previously, for inhomogeneous media consisting of
piecewise homogeneous regions, it is best to seek the solution in
each region first, and then the solutions sewn together by boundary
conditions.  Consequently, surface (boundary) integral equations can
be derived to solve such problems where unknowns only need to be
assigned to the interfaces or boundaries between regions. In this
manner, a 3D problem is reduced to a problem on a 2D manifold,
beating the tyranny of dimensionality.  Moreover, in recent years,
fast algorithms have been developed to solve these surface integral
equations rapidly \cite{Greengard,Coifman,FEACEM}, greatly
underscoring their importance.

To this end, we need to derive the equivalence of the Green's
theorem for vector potential formulation. In the following, we
assume a simple Lorentz gauge so that the equations for homogeneous
region greatly simplify.
In other words, we need to derive Green's
theorem's equivalence for
\begin{equation}\label{5:eq1}
(\nabla^2+k^2)\textbf{A}(\textbf{r})=-\mu \textbf{J}(\textbf{r}).
\end{equation}
where $k^2=\omega^2\mu\varepsilon$ and the time dependence is
$\exp(-i\omega t)$.  It is more expedient to write the above
as\footnote{If $\alpha\ne 1$, the ensuing equation is of the form
\begin{equation}\label{5:eq2a}
\nabla \times \nabla \times \textbf{A}(\textbf{r}) -\alpha^{-1}
\nabla \nabla \cdot \textbf{A}(\textbf{r})-k^2
\textbf{A}(\textbf{r})=\mu \textbf{J}(\textbf{r}).
\end{equation}
But the dyadic Green's function of such an equation can still be
found using methods outlined in \cite{ChewTongHu}. }
\begin{equation}\label{5:eq2}
\nabla \times \nabla \times \textbf{A}(\textbf{r}) - \nabla \nabla
\cdot \textbf{A}(\textbf{r})-k^2 \textbf{A}(\textbf{r})=\mu
\textbf{J}(\textbf{r}).
\end{equation}
We can define a dyadic Green's function that satisfies
\begin{equation}\label{5:eq3}
\nabla \times \nabla \times \overline {\textbf{G}}(\textbf{r},
\textbf{r}') - \nabla \nabla \cdot \overline{\textbf{G}}(\textbf{r},
\textbf{r}')-k^2\overline{\textbf{G}}(\textbf{r},
\textbf{r}')=\overline{\textbf{I}}\delta(\textbf{r}-\textbf{r}').
\end{equation}
The solution to the above is simply
\begin{equation}\label{5:eq4}
\overline{\textbf{G}}(\textbf{r}, \textbf{r}')
=\overline{\textbf{I}} \frac{e^{ik|\textbf{r}-\textbf{r}'|}}{4\pi
|\textbf{r}-\textbf{r}'|} = \overline{\textbf{I}}~ g(\textbf{r},
\textbf{r}').
\end{equation}
From the above, using methods outlined in \cite[Chapter 8]{Chew}, as
well as in Appendix \ref{Appendix:A}, we have for region 1,
\begin{multline}\label{5:eq5}
  \left.
   \begin{aligned}{}
            \textbf{r}\in {V}_1&,&\textbf{A}_1(\textbf{r}) \\
            \textbf{r}\in {V}_2&,&0~~~
   \end{aligned}
\ \right\} = \textbf{A}_{\text{inc}}(\textbf{r})+\int_S dS'
\left\{\mu_1 \overline{\textbf{G}}_1(\textbf{r},\textbf{r}')
\cdot\hat{{n}}'\times \textbf{H}_1(\textbf{r}')-\Big[\nabla' \times
\overline{\textbf{G}}_1(\textbf{r},
\textbf{r}')\Big]\cdot\hat{{n}}'\times \textbf{A}_1(\textbf{r}')\right\}\\
+\int_S dS'~\hat{{n}}'\cdot
\left\{\overline{\textbf{G}}_1(\textbf{r},\textbf{r}') \nabla' \cdot
\textbf{A}_1(\textbf{r}')-\textbf{A}_1(\textbf{r}')\nabla'\cdot
\overline{\textbf{G}}_1(\textbf{r},\textbf{r}')\right\}.
\end{multline}
We can rewrite the above using scalar Green's function as
\begin{multline}\label{5:eq6}
\left.
   \begin{aligned}{}
            \textbf{r}\in {V}_1&,&\textbf{A}_1(\textbf{r}) \\
            \textbf{r}\in {V}_2&,&0~~~
   \end{aligned}
\ \right\}
=\textbf{A}_{\text{inc}}(\textbf{r})+\int_S dS'
\Big\{\mu_1 g_1(\textbf{r},\textbf{r}')\hat{{n}}'\times
\textbf{H}_1(\textbf{r}')-
\nabla' g_1(\textbf{r},\textbf{r}')\times\hat{{n}}'\times \textbf{A}_1(\textbf{r}')\Big\}\\
+\int_S dS'\Big\{-\hat{{n}}'g_1(\textbf{r},\textbf{r}') \nabla'
\cdot
\textbf{A}_1(\textbf{r}')+(\hat{{n}}'\cdot\textbf{A}_1(\textbf{r}'))\nabla'
g_1(\textbf{r},\textbf{r}')\Big\}.
\end{multline}
A similar equation can be derived for region 2. These equations can
be used to formulate surface integral equations for scattering. The
lower parts of the above equations are known as the extinction
theorem \cite{Chew,Ishimaru}.

As a side note, one can use the scalar Green's theorem directly on
\eqref{5:eq1} and obtain
\begin{multline}\label{5:eq6b}
\left.
   \begin{aligned}{}
            \textbf{r}\in {V}_1&,&\textbf{A}_1(\textbf{r}) \\
            \textbf{r}\in {V}_2&,&0~~~
   \end{aligned}
\ \right\}
=\textbf{A}_{\text{inc}}(\textbf{r})-\int_S dS' \Big\{
g_1(\textbf{r},\textbf{r}')\hat{{n}}'\cdot\nabla'\v A_1(\v r')- \hat
n'\cdot\nabla' g_1(\textbf{r},\textbf{r}')\v A_1(\v r')\Big\}.
\end{multline}
After some lengthy manipulations, \eqref{5:eq6} becomes
\eqref{5:eq6b} as shown in Appendix \ref{Appendix:A}.  In the above
derivation, there is a surface integral at infinity that can be
shown to vanish as in \cite[Chapter 8]{Chew} when radiation
condition is invoked.

\section{PEC Scatterer Case}

For a PEC scatterer, we have proved that $\hat n\times \v A_1=0$.
Since $\nabla\cdot \v A_1=i\omega\mu_1\varepsilon_1\Phi$ and that
$\Phi=0$ on a PEC surface.  Hence, for surface sources that satisfy
the PEC scattering solution, the above becomes
\begin{align}\label{eq6a}
\left.
   \begin{aligned}{}
            \textbf{r}\in {V}_1&,&\textbf{A}_1(\textbf{r}) \\
            \textbf{r}\in {V}_2&,&0~~~
   \end{aligned}
\ \right\}
 =\textbf{A}_{\text{inc}}(\textbf{r})+\int_S
dS' \Big\{\mu_1 g_1(\textbf{r},\textbf{r}')\hat{{n}}'\times
\textbf{H}_1(\textbf{r}') +
\hat{{n}}'\cdot\textbf{A}_1(\textbf{r}')\nabla'
g_1(\textbf{r},\textbf{r}')\Big\}.
\end{align}
The first term in the integral comes from the induced surface
current flowing on the PEC surface. It will be interesting to ponder
the meaning of the second term.  It is to be noted that the surface
charge on the PEC surface is given by
\begin{align}
\hat n\cdot\varepsilon_1\v E_1 =\hat n\cdot \varepsilon (i\omega\v
A_1 -\nabla \Phi_1)
\end{align}
The scalar potential $\Phi$ can be obtained from the vector
potential using Lorentz gauge, namely, $\Div \v
A=i\omega\mu\varepsilon\Phi$.  Hence, one can view that $\hat n\cdot
\v A$ as the contribution to the surface charge from the vector
potential $\v A$.  In fact, using the Lorentz gauge, and that $\v
E=i\omega\v A-\nabla\Phi$, one can recover from the above that the
electric field in region 1 outside the PEC is given by (see Appendix
\ref{Appendix:B})
\begin{align}\label{eq48}
\v E_1=\v E_{\rm{inc}}+\int_S dS'\left\{ i\omega\mu_1 g_1(\v r, \v
r') \v J_1(\v r') - \Grad g_1(\v r, \v r')\frac{\sigma_1(\v
r')}{\varepsilon_1(\v r')}\right\}
\end{align}
where $\v J_1=\hat n \times \v H_1$ and $\Div \v
J_1=i\omega\sigma_1$. The above is just the traditional relationship
between the $\v E$ field in region 1 and the sources on the PEC
surface.

We can rewrite \eqref{eq6a} in terms of two integral equations
\begin{eqnarray}\label{eq52}
\textbf{A}_1(\textbf{r}) = \textbf{A}_\text{inc}(\textbf{r}) +
               \int_{S}\,d\,S'\Big\{\mu_1g_1(\textbf{r},\textbf{r}')\textbf{J}_1(\textbf{r}') +
               \Sigma_{1}(\textbf{r}')\nabla'g(\textbf{r},\textbf{r}')\Big\}
\end{eqnarray}
\begin{eqnarray}\label{eq53}
\Sigma_{1}(\textbf{r}) = \Sigma_\text{inc}(\textbf{r}) +
\int_{S}\,d\,S'\Big\{\mu_1g_1(\textbf{r},\textbf{r}')~\hat{n}\cdot\textbf{J}_1(\textbf{r}')
                       + \Sigma_1(\textbf{r}')~\hat{n}\cdot\nabla'g(\textbf{r},\textbf{r}')\Big\}
\end{eqnarray}
where the second equation is obtained by $\hat{n}\cdot$ the first
equation. Also, the boundary condition is such that
$\hat{n}\times\textbf{A}_1(\textbf{r}) = 0$ on $S$. The above can be
solved by the subspace projection method such as the Galerkin's
\cite{Galerkin} or moments methods \cite{Kravchuk,Harrington2}.
The unknowns are $\v J_1$ and $\Sigma_1$ while $\v A_\text{inc}$ and
$\Sigma_\text{inc}$ are known.
We expand the unknowns in terms of basis functions $\v J_n$ and
$\sigma_m$ that span the subspaces of $\v J_1$ and $\Sigma_1$,
respectively.  Namely,
\begin{eqnarray}
\textbf{J}_1(\textbf{r}') =
\sum_{n=1}^{N}j_n\textbf{J}_n(\textbf{r}')
\end{eqnarray}
\begin{eqnarray}
\Sigma_1(\textbf{r}') = \sum_{m=1}^{M}s_m\sigma_m(\textbf{r}')
\end{eqnarray}
We choose $\textbf{J}_n(\textbf{r}')$ to be divergence conforming
tangential current so that the vector potential $\textbf{A}_1$ that
it produces is also divergence conforming \cite{Dai}. In the above,
$\sigma_m(\textbf{r}')$ can be chosen to approximate a surface
charge well. After expanding the unknowns, we project the field that
they produce onto the subspace spanned by the same unknown set as in
the process of testing in the Galerkin's method. Consequently,
\eqref{eq52} and \eqref{eq53} become
\begin{eqnarray}
0 \, = \,
\langle\textbf{J}_{n'}(\textbf{r}),\textbf{A}_\text{inc}(\textbf{r})\rangle
    & + & \mu_1\sum_{n=1}^{N}\langle\textbf{J}_{n'}(\textbf{r}),g_1(\textbf{r},\textbf{r}'),\textbf{J}_n(\textbf{r})\rangle j_n \nonumber \\
    & + & \sum_{m=1}^{M}s_m
    \langle\textbf{J}_{n'}(\textbf{r}),\nabla'g_1(\textbf{r},\textbf{r}'),\sigma_m(\textbf{r}')\rangle\\
\sum_{m=1}^{M}s_m\langle\sigma_{m'}(\textbf{r}),\sigma_m(\textbf{r})\rangle \, & = &\, \langle\sigma_{m'}(\textbf{r}),\Sigma_\text{inc}(\textbf{r})\rangle \nonumber \\
& + & \mu_1\sum_{n=1}^{N}\langle\sigma_{m'}(\textbf{r}),\hat{n}g_1(\textbf{r},\textbf{r}'),\textbf{J}_n(\textbf{r}')\rangle j_n \nonumber \\
& + &
\sum_{m=1}^{M}\langle\sigma_{m'}(\textbf{r}),\hat{n}\cdot\nabla'g_1(\textbf{r},\textbf{r})',\sigma_m(\textbf{r}')\rangle
s_m
\end{eqnarray}
The above is a matrix system of the form
\begin{eqnarray}
0 = \textbf{a}_\text{inc} +
\dyadg{\Gamma}_{1,\textbf{J},\textbf{J}}\cdot\textbf{j} +
\dyadg{\Gamma} _{1,\textbf{J},\sigma}\cdot\textbf{s}
\end{eqnarray}
\begin{eqnarray}
\overline {\textbf{B}}\cdot\textbf{s} = \vg\Sigma_\text{inc} +
\dyadg{\Gamma}_{1,\sigma,\textbf{J}}\cdot\textbf{j} +
\dyadg{\Gamma}_{1,\sigma,\sigma}\cdot\textbf{s}
\end{eqnarray}
where $\v j$ and $\v s$ are unknowns, while $\v a_\text{inc}$ and
$\vg \Sigma_\text{inc}$ are known.  In detail, elements of the above
matrices and vectors are given by
\begin{eqnarray}
[\textbf{a}_\text{inc}]_{n'} = \,
\langle\textbf{J}_{n'}(\textbf{r}),\textbf{A}_\text{inc}(\textbf{r})\rangle
\end{eqnarray}
\begin{eqnarray}
[\dyadg{\Gamma}_{1,\textbf{J},\textbf{J}}]_{n',n} = \mu\langle
\textbf{J}_{n'}(\textbf{r}),g_1(\textbf{r},\textbf{r}'),\textbf{J}_n(\textbf{r}')\rangle
\end{eqnarray}
\begin{eqnarray}
[\dyadg{\Gamma}_{1,\textbf{J},\sigma}]_{n',m} = \langle
\textbf{J}_{n'}(\textbf{r}),\nabla'g_1(\textbf{r},\textbf{r}'),\sigma_m(\textbf{r}')\rangle
\end{eqnarray}
\begin{eqnarray}
[\overline {\textbf{B}}]_{m',m} = \, \langle
\sigma_{m'}(\textbf{r}),\sigma_{m}(\textbf{r})\rangle
\end{eqnarray}
\begin{eqnarray}
[\vg \Sigma _\text{inc}]_{m'} = \,
\langle\sigma_{m'}(\textbf{r}),\Sigma_\text{inc}(\textbf{r})\rangle
\end{eqnarray}
\begin{eqnarray}
[\dyadg{\Gamma}_{1,\sigma,\textbf{J}}]_{m',n} = \mu_1 \langle
\sigma_{m'}(\textbf{r}),\hat{n}~g_1(\textbf{r},\textbf{r}'),\textbf{J}_n(\textbf{r}')\rangle
\end{eqnarray}
\begin{eqnarray}
[\dyadg{\Gamma}_{1,\sigma,\sigma}]_{m',m} = \, \langle
\sigma_{m'}(\textbf{r}),\hat{n}\cdot\nabla'g_1(\textbf{r},\textbf{r}'),\sigma_m(\textbf{r}')\rangle
\end{eqnarray}
\begin{eqnarray}
[\v j]_n = j_n, \quad\quad [\v s]_m = s_m
\end{eqnarray}
Furthermore, in the above,
\begin{eqnarray}
\langle\textbf{f}(\textbf{r}),\textbf{h}(\textbf{r})\rangle =
\int_{S}\,d\,S\textbf{f}(\textbf{r})\cdot\textbf{h}(\textbf{r})
\end{eqnarray}
\begin{eqnarray}
\langle\textbf{f}(\textbf{r}),\gamma(\textbf{r},\textbf{r}'),\textbf{h}(\textbf{r})\rangle
=
                     \int_{S}\,d\,S\textbf{f}(\textbf{r})\cdot\int_{S}\,d\,S'\gamma(\textbf{r},\textbf{r}')\textbf{h}(\textbf{r}')
\end{eqnarray}
where $\textbf{f}(\textbf{r})$ and $\textbf{h}(\textbf{r})$ can be
replaced by scalar functions, and $\gamma(\textbf{r},\textbf{r}')$
can be replaced by a vector function with the appropriate inner
products between them.

The $\dyadg \Gamma$ matrices above are different matrix
representations of the scalar Green's function and its derivative.
It is to be noted that all the $\dyadg \Gamma$ matrices above do not
have low-frequency catastrophe as in the matrix representation of
the dyadic Green's function.  Hence, the above behaves like the
augmented electric field integral equation (A-EFIE) \cite{Qian}.

\section{Vector Potential Plane Wave}

A time-harmonic vector potential plane wave is the solution to the
equation
\begin{align}
\left( \nabla^2+k^2\right) \v A =0
\end{align}
But it seems odd that $A_x$, $A_y$, and $A_z$ are decoupled from
each other.   To dispel this notion, we should think of $\v A$ as
the solution to
\begin{align}
\left( \nabla^2+k^2\right) \v A =-\mu \v J
\end{align}
The vector potential above satisfies the Lorentz gauge via the
charge continuity equation.  By taking the divergence of the above,
we have
\begin{align}
\left( \nabla^2+k^2\right)\Div \v A =-\mu\Div \v J=-i\omega\mu \rho
\end{align}
where $\Div \v A=i\omega\mu\varepsilon\Phi$.

If $\v J$ is due to a Hertzian dipole source
\begin{align}
\v J(\v r)=I \ell \hat \ell \delta(\v r)
\end{align}
the corresponding vector potential $\v A$ is
\begin{align}
\v A(\v r)=\mu I \ell\hat \ell \frac{e^{ikr}}{4\pi r}
\end{align}
We can produce a locally plane wave by letting $\v r=\v r_0+\v s$
where $|\v r_0|\gg |\v s|$.  Then the above spherically wave can be
approximated by a locally plane wave:
\begin{align}
\v A(\v r)\approx \mu I \ell\hat \ell \frac{e^{ikr_0}}{4\pi
r_0}e^{i\v k_0\cdot \v s}=\v a e^{i\v k_0\cdot\v s}
\end{align}
where $\v k_0=k \hat r_0$ and $\hat r_0$ is a unit vector that
points in the direction of $\v r_0$.  It is seen that the components
of $\v A$ generated this way satisfies the gauge condition and are
not independent of each other.  We have to keep this notion in mind
when we generate a vector potential plane wave.

Hence, for a plane wave incident,
\begin{equation}
    {\bf A}_\text{inc}({\bf r})=({\bf a}_\perp+{\bf a}_{k_i})e^{i{\bf k}_i\cdot{\bf r}}
\end{equation}
where ${\bf a}_{k_i}=a_0\hat{k}_i$, and $\hat{k}_i\cdot{\bf
a}_\perp=0$. Therefore
\begin{equation}
    \nabla\cdot{\bf A}_\text{inc}=i{\bf k}_i\cdot{\bf a}_{k_i}e^{i{\bf k}_i\cdot{\bf r}}=ia_0k_ie^{i{\bf k}_i\cdot{\bf r}}
\end{equation}
\begin{equation}
    {\bf B}_\text{inc}=\nabla\times{\bf A}_\text{inc}=i{\bf k}_i\times{\bf A}_\text{inc}=i{\bf k}_i\times{\bf a}_\perp e^{i{\bf k}_i\cdot{\bf r}}
\end{equation}

and
\begin{align}
    {\bf E}_\text{inc}&=\frac{\nabla\times{\bf B}_\text{inc}}{-i\omega\mu\varepsilon}=\frac{i{\bf k}_i\times({\bf k}_i\times{\bf a}_\perp)e^{i{\bf k}_i\cdot{\bf r}}}{\omega\mu\varepsilon}\\
    &=i[k_i^2{\bf a}_\perp-{\bf k}_i({\bf k}_i\cdot{\bf a}_\perp)]\frac{e^{i{\bf k}_i\cdot{\bf r}}}{\omega\mu\varepsilon} \notag \\
    &=i\omega\left[\overline{\bf I}-\hat{k}_i\hat{k}_i\right]\cdot{\bf a}_\perp e^{i{\bf k}_i\cdot{\bf r}}
\end{align}
It is to be noted that if $\v A_\text{inc}$ has only the
longitudinal component, then both $\v E$ and $\v B$ are zero even
though $\v A$ is not zero.  This could happen to leading order along
the axial direction of a Hertzian dipole.


\appendix
\numberwithin{equation}{section}

\section{Derivations of \eqref{5:eq5}, \eqref{5:eq6}, and
\eqref{5:eq6b}}\label{Appendix:A}

\setcounter{equation}{0}


We will ignore the source term $\v J$ in order to derive some
identities similar to Green's theorem.  We begin with the following
equations:
\begin{equation}\label{a:e1}
\nabla  \times \nabla  \times {\bf{A}} - \nabla \nabla  \cdot
{\bf{A}} - {k^2}{\bf{A}} = 0
\end{equation}
\begin{equation}\label{a:e2}
\nabla  \times \nabla  \times {\bf{\overline G}} - \nabla \nabla
\cdot {\bf{\overline G}} - {k^2}{\bf{\overline G}} = {\bf{\bar
I}}\delta \left( {{\bf{r}} - {\bf{r}}'} \right)
\end{equation}
In the above, $\v A=\v A(\v r)$ and $\dyad G=\dyad G(\v r, \v r')$,
but we will suppress these dependent variables for the time being in
the following. First, we dot-multiply \eqref{a:e1} from the right by
$\dyad G\cdot \v a$ where $\v a$ is an arbitrary vector, and then
dot-multiply \eqref{a:e2} from the left by $\v A$ and the right by
$\v a$.  We take their difference, and ignoring the $\nabla\nabla$
term for the time being, to get
\begin{equation}\label{a:e3}
\begin{array}{l}
~~~\nabla  \times \nabla  \times {\bf{A}} \cdot {\bf{\overline G}} \cdot {\bf{a}} - {\bf{A}} \cdot \nabla  \times \nabla  \times {\bf{\overline G}} \cdot {\bf{a}}\\
 = \nabla  \cdot \left( {\nabla  \times {\bf{A}} \times {\bf{\overline G}} \cdot {\bf{a}} + {\bf{A}} \times \nabla  \times {\bf{\overline G}} \cdot {\bf{a}}} \right)
\end{array}
\end{equation}
Integrating \RHS\ of the above over $V$, we have
\begin{equation}\label{a:e4}
\begin{array}{*{20}{l}}
{{{\bf{I}}_1}\cdot{\bf{a}} = \int\limits_S {\hat n\cdot\left( {\nabla  \times {\bf{A}} \times {\bf{\overline G}}\cdot{\bf{a}} + {\bf{A}} \times \nabla  \times {\bf{\overline G}}\cdot{\bf{a}}} \right)dS} }\\
~~~~~~~{ = \int\limits_S {\left[ {\hat n \times \left( {\nabla
\times {\bf{A}}} \right)\cdot{\bf{\overline G}}\cdot{\bf{a}} +
\left( {\hat n \times {\bf{A}}} \right)\cdot\nabla  \times {\bf{\bar
G}}\cdot{\bf{a}}} \right]dS} }
\end{array}
\end{equation}
Including now the $\nabla\nabla$ term gives
\begin{equation}\label{a:e5}
\begin{array}{l}
~~- \nabla \nabla  \cdot {\bf{A}} \cdot {\bf{\overline G}} \cdot {\bf{a}} + {\bf{A}} \cdot \nabla \nabla  \cdot {\bf{\overline G}} \cdot {\bf{a}}\\
= \nabla  \cdot \left( { - \nabla  \cdot {\bf{A~\overline G}} \cdot
{\bf{a}} + {\bf{A}} \cdot \nabla  \cdot {\bf{\overline G}} \cdot
{\bf{a}}} \right)
\end{array}
\end{equation}
Integrating the \RHS\ of the above over $V$, we have
\begin{equation}\label{a:e6}
{{\bf{I}}_2} \cdot {\bf{a}} = \int\limits_S {\hat n \cdot \left( { -
\nabla  \cdot {\bf{A~\overline G}} \cdot {\bf{a}} + {\bf{A}} \cdot
\nabla \cdot {\bf{\overline G}} \cdot {\bf{a}}} \right)dS}
\end{equation}
Letting ${\bf{\overline G}} = g{\bf{\bar I}}$, where $g=g(\v r,\v
r')$, the scalar Green's function, the above becomes
\begin{equation}\label{a:e7}
\begin{array}{l}
{{\bf{I}}_1} \cdot {\bf{a}} = \int\limits_S {\left[ {\hat n \times \nabla  \times {\bf{A}}g \cdot {\bf{a}}{\rm{ + }}\hat n \times {\bf{A}} \cdot \nabla g \times {\bf{a}}} \right]dS} \\
~~~~~~= \int\limits_S {\left[ {\hat n \times \left( {\nabla  \times
{\bf{A}}} \right)g{\rm{ + }}\left( {\hat n \times {\bf{A}}} \right)
\times \nabla g} \right]dS \cdot {\bf{a}}}
\end{array}
\end{equation}
or
\begin{equation}\label{a:e8}
{{\bf{I}}_1} = \int\limits_S {\left[ {\hat n \times \left( {\nabla
\times {\bf{A}}} \right)g{\rm{ + }}\left( {\hat n \times {\bf{A}}}
\right) \times \nabla g} \right]dS}
\end{equation}
Similarly, we have
\begin{equation}\label{a:e9}
{{\bf{I}}_2} = \int\limits_S {\left[ { - \hat n\left( {\nabla  \cdot
{\bf{A}}} \right)g{\rm{ + }}\hat n \cdot {\bf{A}}~\nabla g}
\right]dS}
\end{equation}
Using the above, we get \eqref{5:eq6}.

To get \eqref{5:eq6b}, more manipulations are needed. Using $\hat n
\times \left( {\nabla \times {\bf{A}}} \right) = \left( {\nabla
{\bf{A}}} \right) \cdot \hat n - \left( {\hat n \cdot \nabla }
\right){\bf{A}}$, $\nabla g \times \left( {\hat n \times {\bf{A}}}
\right) = \hat n\left( {{\bf{A}} \cdot \nabla g} \right) - \left(
{\hat n \cdot \nabla g} \right){\bf{A}}$
\begin{equation}\label{a:e10}
{{\bf{I}}_1} = \int\limits_S {\left[ { - \left( {\hat n \cdot \nabla
{\bf{A}}} \right)g{\rm{ + }}\left( {\nabla {\bf{A}}} \right) \cdot
\hat ng - \hat n\left( {{\bf{A}} \cdot \nabla g} \right) + \left(
{\hat n \cdot \nabla g} \right){\bf{A}}} \right]dS}
\end{equation}
First, we look at
\begin{eqnarray}\label{a:e11}
{\bf{I}}_3 & = & \int\limits_S\,\hat{n}[-\nabla\cdot\bar{\bf{A}}\,~g
- \bar{\bf{A}}\cdot\nabla\,g]d\,S
\nonumber \\
          & = & \int\limits_V\,d\,V~\nabla(-\nabla\cdot\bar{\bf{A}}\,~g - \bar{\bf{A}}\cdot\nabla\,g)
\nonumber \\
          & = & \int\limits_V\,d\,V~[-\nabla\nabla\cdot\bar{\bf{A}}\,~g - \bar{\bf{A}}\cdot\nabla\nabla\,g
          - \nabla\bar{\bf{A}}\cdot\,\nabla g -
          \nabla\cdot\bar{\bf{A}}~\nabla\,g]\\
\label{a:e12}
{\bf{I}}_4 & = & \int\limits_S\,d\,S~[g~\nabla{\bf{A}}\cdot\hat{n} + \hat{n}\cdot{\bf{A}}~\nabla\,g] \nonumber \\
          & = & \int\limits_V\,d\,V\,~\nabla\cdot[(g\nabla{\bf{A}})^{t} + {\bf{A}}~\nabla\,g]
\end{eqnarray}
Furthermore, with the knowledge that
\begin{eqnarray}
\nabla\cdot[(g\nabla {\bf{A}})^{t} -  {\bf{A}}\nabla\,g] & = & \partial_i[g\partial_k {\bf{A}}_i + {\bf{A}}_i \partial_jg] \\
& = & (\partial_ig)\partial_k {\bf{A}}_i + g\partial_i\partial_k {\bf{A}}_i + \partial_i {\bf{A}}_i\partial_jg + {\bf{A}}_i\partial_i\partial_jg \\
& = & \nabla {\bf{A}}\cdot\nabla\,g + g~\nabla\nabla\cdot {\bf{A}} +
\nabla\cdot {\bf{A}}~\nabla\,g + {\bf{A}}\cdot\nabla\nabla\,g
\end{eqnarray}
it is seen that ${\bf{I}}_3+{\bf{I}}_4=0$. Using this fact, we can
show \eqref{5:eq6b}, or that
\begin{eqnarray}
{\bf{I}}_1 + {\bf{I}}_2 =
\int\limits_S\,d\,S~[(\hat{n}\cdot\nabla\,g){\bf{A}} -
g~\hat{n}\cdot\nabla{\bf{A}}]
\end{eqnarray}

\section{Derivation of \eqref{eq48}}\label{Appendix:B}

\begin{eqnarray}\label{b:eq1}
\textbf{A}_1 = \textbf{A}_\text{inc} +
\int_{S}\,d\,S'\Big\{\mu_1g_1\textbf{J}_1 +
(\hat{n}'\cdot\textbf{A}_1)\nabla'g_1\Big\} \label{eq1}
\end{eqnarray}
\begin{eqnarray}
\nabla\cdot\textbf{A}_1 = \nabla\cdot\textbf{A}_\text{inc} +
\int_{S}\,d\,S'\Big\{\mu_1g_1\nabla'\cdot\textbf{J}_1 -
                                                  (\hat{n}'\cdot\textbf{A}_1)\nabla^{2}g_1\Big\}
\label{b:eq2}
\end{eqnarray}
Since $\nabla\cdot\textbf{A} = i\omega\mu\varepsilon\Phi$, the above
becomes
\begin{eqnarray}
i\omega\mu_1\varepsilon_1\Phi_1 =
i\omega\mu_1\varepsilon_1\Phi_\text{inc} +
\int_{S}\,d\,S'\Big\{\mu_1g_1i\omega\sigma_1 +
(\hat{n}'\cdot\textbf{A}_1)k_1^2g_1\Big\} \label{b:eq3}
\end{eqnarray}
or
\begin{eqnarray}
\Phi_1 = \Phi_\text{inc} +
\int_{S}\,d\,S'\bigg\{g_1\frac{\sigma_1}{\varepsilon_1} -
i\omega(\hat{n}'\cdot\textbf{A}_1)g_1\bigg\} \label{b:eq4}
\end{eqnarray}
Since $\textbf{E} = i\omega\textbf{A} - \nabla\Phi$, using
\eqref{b:eq1} and \eqref{b:eq4}, we have
\begin{eqnarray}
\textbf{E}_1 = \textbf{E}_\text{inc} +
\int_{S}\,d\,S'\bigg\{i\omega\mu_1g_1\textbf{J}_1 - \nabla
g_1\frac{\sigma_1}{\varepsilon_1}
                                             + i\omega(\hat{n}'\cdot\textbf{A}_1)\nabla'g_1 + i\omega(\hat{n}'\cdot\textbf{A}_1)\nabla g_1\bigg\}
\label{b:eq5}
\end{eqnarray}
The last two terms cancel each other.

\subsection*{Acknowledgements} This work was supported in part by the
USA NSF CCF Award 1218552, SRC Award 2012-IN-2347, at the University
of Illinois at Urbana-Champaign, by the Research Grants Council of
Hong Kong (GRF 711609, 711508, and 711511), and by the University
Grants Council of Hong Kong (Contract No. AoE/P-04/08) at HKU. The
author is grateful to M. Wei, H. Gan, C. Ryu, T. Xia, Y. Li, Q. Liu,
and L. Meng for helping to typeset the manuscript.


\end{document}